\documentclass[twocolumn]{revtex4-1} 
\usepackage{graphicx}
\usepackage{amssymb} 
\usepackage{amsmath}
\usepackage{color}
\usepackage[normalem]{ulem}

\newcommand{\Pm}{{$^{147}$Pm}}
\newcommand{\Tr}{{$^3$H}}
\begin{document} 

\title{Analysis of the attainable efficiency of a direct-bandgap betavoltaic element}

\author{A.V.~Sachenko}
\affiliation{V. Lashkaryov Institute of Semiconductor Physics, NAS of Ukraine, 41 prospect Nauky, 03028 Kyiv, Ukraine }
\author{A.I.~Shkrebtii}
\affiliation{University of Ontario Institute of Technology, 2000 Simcoe Street North, Oshawa, ON,  L1H 7K4 Canada}
\author{R.M.~Korkishko}
\affiliation{V. Lashkaryov Institute of Semiconductor Physics, NAS of Ukraine, 41 prospect Nauky, 03028 Kyiv, Ukraine }
\author{V.P.~Kostylyov}
\affiliation{V. Lashkaryov Institute of Semiconductor Physics, NAS of Ukraine, 41 prospect Nauky, 03028 Kyiv, Ukraine }
\author{ N.P.~Kulish}
\affiliation{V. Lashkaryov Institute of Semiconductor Physics, NAS of Ukraine, 41 prospect Nauky, 03028 Kyiv, Ukraine }
\author{ I.O.~Sokolovskiy}
\affiliation{V. Lashkaryov Institute of Semiconductor Physics, NAS of Ukraine, 41 prospect Nauky, 03028 Kyiv, Ukraine }
\author{ M.~Evstigneev}
\email[Corresponding author: ]{mevstigneev@mun.ca}
\affiliation{Department of Physics and Physical Oceanography, Memorial University of Newfoundland, St. John's, NL, A1B 3X7  Canada}

\begin{abstract}
Conversion of energy of beta-particles into electric energy in a p-n junction based on direct-bandgap semiconductors, such as GaAs, considering realistic semiconductor system parameters is analyzed. An expression for the collection coefficient, $Q$, of the electron-hole pairs generated by beta-electrons is derived taking into account the existence of the dead layer. We show that the collection coefficient of beta-electrons emitted by a \Tr-source to a GaAs p-n junction is close to 1 in a broad range of electron lifetimes in the junction, ranging from $10^{-9}$ to $10^{-7}$ s. For the combination \Pm/GaAs, $Q$ is relatively large ($\ge 0.4$) only for quite long lifetimes (about $10^{-7}$ s) and large thicknesses (about $100\,\mu$m) of GaAs p-n junctions. For realistic lifetimes of minority carriers and their diffusion coefficients, the open-circuit voltage realized due to the irradiation of a GaAs p-n junction by beta-particles is obtained. The attainable beta-conversion efficiency $\eta$ in the case of a \Tr/GaAs combination is found to exceed that of the \Pm/GaAs combination.
\end{abstract}

\pacs{xxx}

\maketitle 
\section{Introduction}
Betavoltaic effect refers to the electric power production by a p-n junction bombarded by beta-particles that ionize the semiconductor material. Among the advantages of beta-batteries are their long service duration, amounting to years or even decades, and the possibility to use in the hard-to-reach areas. Betavoltaics and photovoltaics are related disciplines. In both cases, electric power results from the separation of electron-hole pairs produced by beta-electrons or photons by a p-n junction in the presence of a load in the circuit. In comparison to photovoltaics, publications in the field of the basic principles and applications of betavoltaic elements have been less numerous initially (see, e.g., Refs.~\cite{Rap54, Pfa54, Rap56, Fli64, Ols73, Ols74, Olstech}), but started to attract the attention of the researchers in the recent years \cite{And00, Bow02, Ada12, Ols12}. 

The main task in betavoltaic design is the choice of a beta-source/semiconductor combination, which should meet certain requirements. In particular, the beta-particles produced by the source must be absorbed efficiently by the semiconductor. Within the semiconductor, the diffusion length of the electron-hole pairs generated by the beta-flux should be large enough to allow them to reach the p-n junction with as little losses as possible. Because only the relatively low-energy beta-electrons are utilized effectively (with energies varying between 5 and 70 keV) for the realistic semiconductor thicknesses, three main beta-sources are presently employed in betavoltaic applications: Tritium \Tr, Nickel $^{63}$Ni, and Promethium \Pm. The respective mean energies of the electrons produced by these sources are 5.7, 18, and 62 keV.

The efficiency, $\eta$, of a betavoltaic converter is proportional to the collection coefficient, $Q$, of the electron-hole pairs generated by the beta-flux.  In Refs.~\cite{Pfa54, Olstech}, $Q$ was calculated under the assumption that the generation function of electron-hole pairs by a beta-flux $g(x) \propto \exp(-\alpha x)$. In reality, the generation function is close to zero within the so-called ``dead layer'' under the front surface, and exhibits a maximum at some distance $x_m$ from the surface \cite{Dmi78}. This implies that this exponential approximation is correct starting from some $x$-value greater than $x_m$. The emergence of the maximum in the $g(x)$ curve is due to the fact that, initially, the primary electrons pass through the semiconductor with only weak scattering. The dead layer thickness $x_m$ increases with the energy of the incident beta-electrons. For GaAs, $x_m$ is in the range 0.1 -- 1 $\mu$m \cite{Dmi78}.

Although the works \cite{Pfa54, Olstech} do report analytical expressions for $Q$ (obtained under the assumption of the absence of the dead layer), the values of $Q = 1$ and 0.7 were used in the calculations of beta-conversion efficiency \cite{Olstech, Ols12}. While the value $Q = 1$ corresponds to the limiting conversion efficiency that is maximal in principle, the choice $Q = 0.7$ was not explained in \cite{Olstech, Ols12}.

In this work, we derive an expression for $Q$ taking the dead layer into account, and also using the realistic values of the nonradiative Shockley-Reed-Hall (SRH) recombination lifetime, $\tau_{SR}$, for direct-bandgap semiconductors. In such materials, the values of $\tau_{SR}$ are usually short, and are in the range of $10^{-9}-10^{-7}$ s. We use the so obtained collection coefficient to derive the expression for the realistically attainable beta-conversion efficiency $\eta$ of various combinations of beta-sources and direct-bandgap semiconductors. When calculating the efficiency, we focus on GaAs as a typical example. We show that decreasing $\tau_{SR}$ and increasing the dead layer thickness leads to a strong reduction of $Q$ below 1, and to the corresponding reduction of the beta-conversion efficiency.

\section{Analysis of the collection coefficient}
\begin{figure}[t!] 
\includegraphics[scale=0.4]{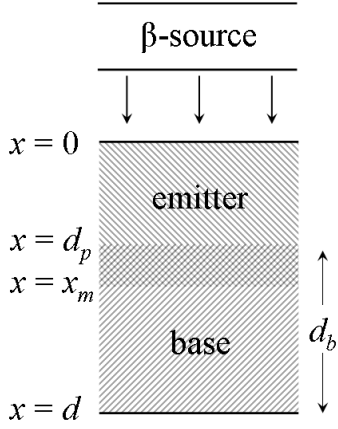}
\caption{Schematic illustration of a p-n junction of thickness $d = d_p + d_b$, where $d_p$ is emitter depth and $d_b$ is base thickness. The dead layer of thickness $x_m$ extends into the base region.}
\label{fig1}
\end{figure}

We assume that the electron-hole pairs are generated only weakly within the dead layer, $x < x_m$, while for $x > x_m$, the generation function has the form $g(x) = I_0\,\exp(-\alpha (x-x_m))$, where $I_0$ is the electron-hole pair generation rate in the $x_m$-plane, and $\alpha^{-1}$ is the characteristic decay length. Furthermore, we assume that $d_p < x_m$ and $S_d \ll D/L$, $d_p$ being the junction depth, $S_d$ the recombination rate on the back surface of the base, and $L$ and $D$ the diffusion length and coefficient of the excess electron-hole pairs generated in the base region. The sketch of our structure is summarized in Fig.~\ref{fig1}.

Apart from the SRH mechanism with the lifiteme $\tau_{SR}$, the electron-hole pairs in GaAs also recombine radiatively; the characteristic time of this process is $\tau_r = (AN_d)^{-1}$, where $A$ is the radiative recombination coefficient, and $N_d$ is the base doping concentration. Therefore, the diffusion length can be written as
\begin{equation}
L = (D\tau_b)^{1/2}\ ,
\label{1}
\end{equation}
with $\tau_b = \left(\tau_{SR}^{-1} + \tau_r^{-1}\right)^{-1}$ being the effective lifiteme in the neutral base region.

Continuity equation for the excess concentration of the electron-hole pairs, $\Delta p_1$, within the dead layer (i.e., for $x < x_m$, region 1), where generation is negligible, has the form
\begin{equation}
\frac{d^2\Delta p_1}{dx^2} - \frac{\Delta p_1}{L^2} = 0\ ,
\label{2}
\end{equation}
In the rest of the semiconductor ($x > x_m$, region 2), the continuity equation for the excess electron-hole pair density, $\Delta p_2$, is
\begin{equation}
\frac{d^2\Delta p_2}{dx^2} - \frac{\Delta p_2}{L^2} = -\frac{\alpha I_0\,e^{-\alpha (x-x_m)}}{D}\ .
\label{3}
\end{equation}
The equations (\ref{2}) and (\ref{3}) are supplemented by the boundary conditions
\begin{eqnarray}
&&\Delta p_1(x = d_p) = 0\ ,\ \ \frac{d\Delta p_2}{dx}(x = d) = 0\ , \nonumber \\ 
&&\Delta p_1(x = x_m) = \Delta p_2(x = x_m)\  ,\nonumber \\
&&\frac{d\Delta p_1}{dx}(x = x_m) = \frac{d\Delta p_2}{dx}(x = x_m)\ .
\label{4}
\end{eqnarray}
The first condition reflects the fact that the electron-hole pairs are separated at the junction depth. The second one indicates the absence of surface recombination at the back of the base. The remaining two expressions are the usual continuity conditions for $\Delta p(x)$ and $d\Delta p(x)/dx$ at $x = x_m$. The collection coefficient is then defined as the ratio of the current at the junction depth, $d_p$, to the pair generation rate in the plane of highest generation at $x = x_m$:
\begin{equation}
Q = \frac{D}{I_0}\frac{d\Delta p_1}{dx}(x = d_p)\ .
\label{5}
\end{equation}
The solution of (\ref{2}) and (\ref{3}) that satisfies the first two conditions (\ref{4}) can be written as
\begin{eqnarray}
&&\Delta p_1(x) = C\sinh\frac{x - d_p}{L}\ ,\nonumber \\
&&\Delta p_2(x) = C'\cosh\frac{x - d}{L} \nonumber \\
&&\ \ \ \ \ \ \ \ \ \ + B\left(e^{-\alpha(x - x_m)} - \beta\,e^{-x/L}\right)\ ,\nonumber \\
&&B = \frac{\alpha\,I_0\,L^2}{D\left(1 - \alpha^2L^2\right)}\ ,\ \ \beta = \alpha L\exp{\left[\left(\frac{1}{L} - \alpha\right)d\right]}
\end{eqnarray}
with constants $C$, $C'$ to be determined from the remaining two conditions (\ref{4}). This procedure yields:
\begin{eqnarray}
&&Q = \alpha L\,\times \nonumber \\
&&\frac{\alpha L\left(\cosh\frac{d - x_m}{L} - e^{-\alpha(d - x_m)}\right) -\sinh\frac{d - x_m}{L}}{\left[(\alpha L)^2-1\right]\cosh\frac{d - d_p}{L}}\ .
\label{8}
\end{eqnarray}
If $d - x_m \gg L$ and $\alpha(d - x_m) \gg 1$, this expression simplifies to
\begin{equation}
Q = \frac{\alpha L}{1 + \alpha L}e^{(d_p - x_m)/L}\ .
\label{8a}
\end{equation}

\begin{figure}[t!] 
\includegraphics[scale=0.25]{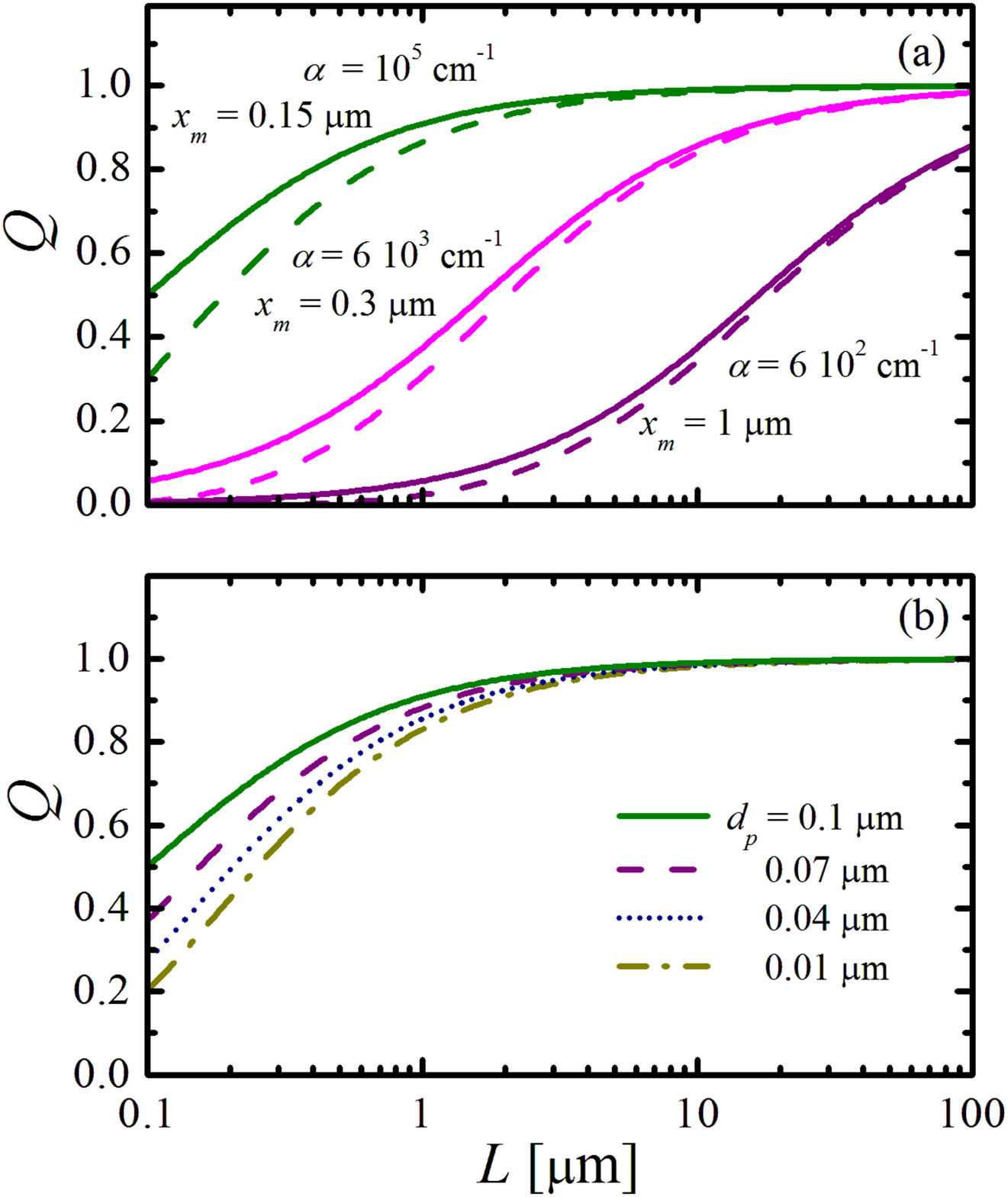}
\caption{(a) Collection coefficient, $Q$, as a function of the diffusion length, $L$,  for different absorption coefficients, $\alpha$, in the limit $\alpha (d - x_m) \gg 1$, $d - x_m \gg L$, see Eq.~(\ref{8a}). The values used, $\alpha = 10^5, 6\cdot 10^3$, and $6\cdot 10^2$\,cm$^{-1}$, approximately correspond to the respective mean beta-energies of $5.7, 20$, and $60$\,keV for GaAs-based p-n junction \cite{Tri67}. The dashed curves are calculated for different dead layer thicknesses, $x_m$, and $d_p = 10^{-5}$ cm. The solid curves are from the standard relation $Q = \alpha L/(1 + \alpha L)$, valid in the absence of the dead layer. (b) Collection coefficient (\ref{8a}) for different junction depth values for $x_m = 10^{-5}$\,cm and $\alpha = 10^5$\,cm$^{-1}$, corresponding to the beta-particle energy of about 5.7\,eV in the \Tr/GaAs combination.}
\label{fig2}
\end{figure}

Fig.~\ref{fig2} shows the dependence of the collection coefficient $Q$ on the diffusion length from Eq.~(\ref{8a}). As seen in this figure, the strongest reduction of $Q$ due to the presence of the dead layer is for the case of the \Tr\  beta-source. The smallest discrepancy in the $Q$-values obtained with and without taking into account the dead layer  is found for the curves corresponding to $\alpha = 6\cdot 10^2$\,cm$^{-1}$, realized in the case of the \Pm-source. In this case, to obtain $Q > 1/2$, one would need the diffusion length $L > 35\ \mu$m. The values $Q \approx 1$ can be achieved only in Si p-n junctions with long minority carrier lifetimes \cite{Gor00}.

In Fig.~\ref{fig2}(b), the junction depth was varied at a fixed electron energy (and thus constant $\alpha$) and dead layer thickness. As seen in this figure, the collection coefficient increases not only upon increasing $L$, but also upon approaching the junction depth to the $x_m$-value. This effect is especially important for small diffusion length $L$.

A further conclusion from Fig.~\ref{fig2} is that collection of the electron-hole pairs generated by the electron flux will be quite efficient in the case when the diffusion length exceeds the dead layer thickness, $L > x_m$. An alternative way to increase $Q$ is to use deeper junctions with $d_p \approx x_m$.

Let us find the relation between the diffusion length and SHR lifetime $\tau_{SR}$ for the case of GaAs. The radiative recombination coefficient $A$ in GaAs is an effective parameter defined by the relation $A = A_0(1 - \gamma_r)$ \cite{Din11}, where $A_0 \approx 6\cdot 10^{-10}$ cm$^3$/s \cite{Sach14}, and $\gamma_r$ is the photon re-absorption coefficient. In our calculations, we assumed the value $A = 2\cdot 10^{-10}$\,cm$^3$/s, as can be derived for poorly reflecting GaAs-based plane-parallel p-n structures without multiple reflection using the approach from \cite{Din11}. In the work \cite{Sach14}, it was shown that for realistic lifetimes  $\tau_{SR}$, the open-circuit voltage $V_{OC}$ of GaAs-based p-n junctions increases with the base doping level, $N_d$, and, taking into account the interband Auger recombination, it has a maximum at $N_d \approx 10^{17}$\,cm$^{-3}$.

Let us first assume that the GaAs p-n junction base is of p-type, and the diffusion coefficient of electron-hole pairs is 50 cm$^2$/s. Then, for $A \approx 2\cdot 10^{-10}$\,cm$^3$/s, $N_d = 10^{17}$ cm$^{-3}$, and lifetimes $\tau_{SR} = 10^{-9}, 10^{-8}$, and $10^{-7}$ s, diffusion length $L$ has the respective values of 2.2, 6.45, and 12.9 $\mu$m. 

\begin{figure}[t!] 
\includegraphics[scale=0.25]{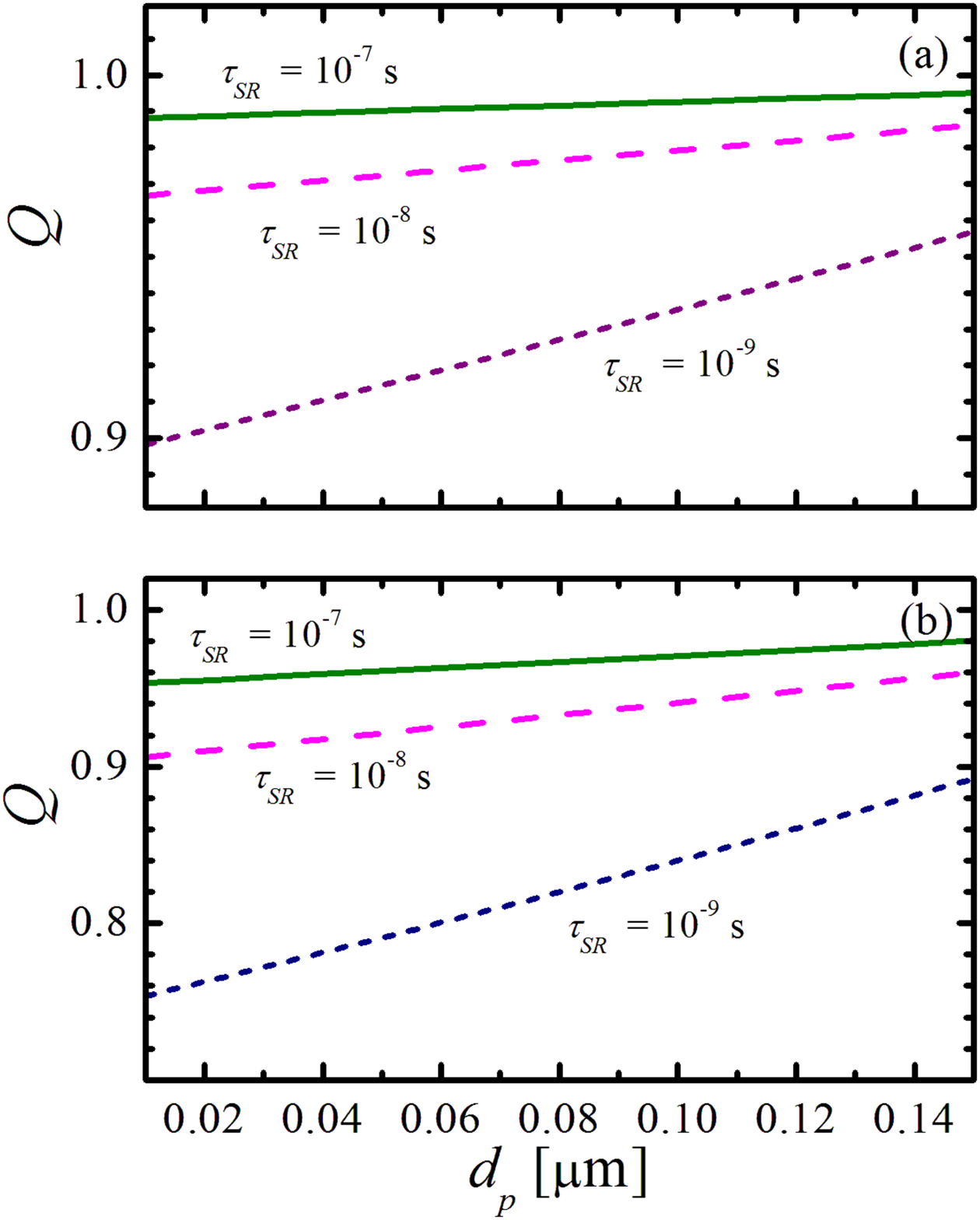}
\caption{Collection coefficient $Q$ of a \Tr/GaAs betavoltaic pair as a function of the junction depth for (a) p-type base and (b) n-type base.}
\label{fig3}
\end{figure}

Fig.~\ref{fig3}(a) shows the dependence of the collection coefficient, $Q$, of a pair \Tr/GaAs as a function of the junction depth, $d_p$, for these three values of $\tau_{SR}$ at $x_m = 0.15\,\mu$m \cite{Dmi78} and junction thickness $d = 10\,\mu$m. As seen in this figure, $Q$ is close to 1 for $\Delta x = x_m - d_p < 0.1\,\mu$m. For $\Delta x > 0.1\,\mu$m, the $Q$-value decreases with $\Delta x$, but remains rather large.

Presented in Fig.~\ref{fig3}(b) is the collection coefficient vs. $d_p$ for the case when the base region of the p-n junction is of the n-type. In this case, for $\tau_{SR} = 10^{-9}, 10^{-8}$, and $10^{-7}$\,s, and $A \approx 2\cdot 10^{-10}$\,cm$^3$/s and $N_d = 10^{17}$\,cm$^{-3}$, and taking into account that $D = 7$\,cm$^2$/s, the diffusion length $L = 0.83, 2.41$, and $4.83\,\mu$m, respectively. As seen in the figure, in this case $Q$ is also quite large. For $\tau_{SR} = 10^{-7}$ and $10^{-8}$\,s, $Q$ is still close to 1, while for $\tau_{SR} = 10^{-9}$\,s, $Q$ exceeds 0.75 even for small $\Delta x$.

It should be noted that, because of rather strong absorption of the electrons emitted by the \Tr-source by the auxiliary layers of a betavoltaic element (such as protection coating or contact layers), additional reduction of the beta-generated current can take place, leading to the efficiency reduction.

Let us now analyze the collection coefficient for the \Pm/GaAs pair. In this case, according to \cite{Tri67}, $\alpha \approx 600$\,cm$^{-1}$, i.e., excess electron-hole density decays much more slowly than in the \Tr/GaAs case. For this the inequality $\alpha L \gg 1$ is alway satisfied even for the shortest lifeteme of $10^{-9}$\,s. In contrast, for \Pm\ source, $\alpha L = 1.5$ for $L = 25\,\mu$m, while $\alpha L = 0.06$ for $L = 1\,\mu$m, so that $Q$ is always notably smaller than 1.

But this is not the only reason for the reduction of $Q$ in realistic \Pm/GaAs structures. When manufacturing solar cells based on the direct-bandgap semiconductors, such as GaAs, full thicknesses of p-n junctions are chosen rather small (of the order of a few $\mu$m). Such structures were used in \cite{And00}. In contrast, for the \Pm/GaAs pair used in betavoltaics, the situation might be very different, especially for large values of $L$. In this case, the product $\alpha d$ will be small, so that for full absorption of beta-flux much thicker p-n junctions are required compared to those typically used in photovoltaics.

\begin{figure}[t!] 
\includegraphics[scale=0.25]{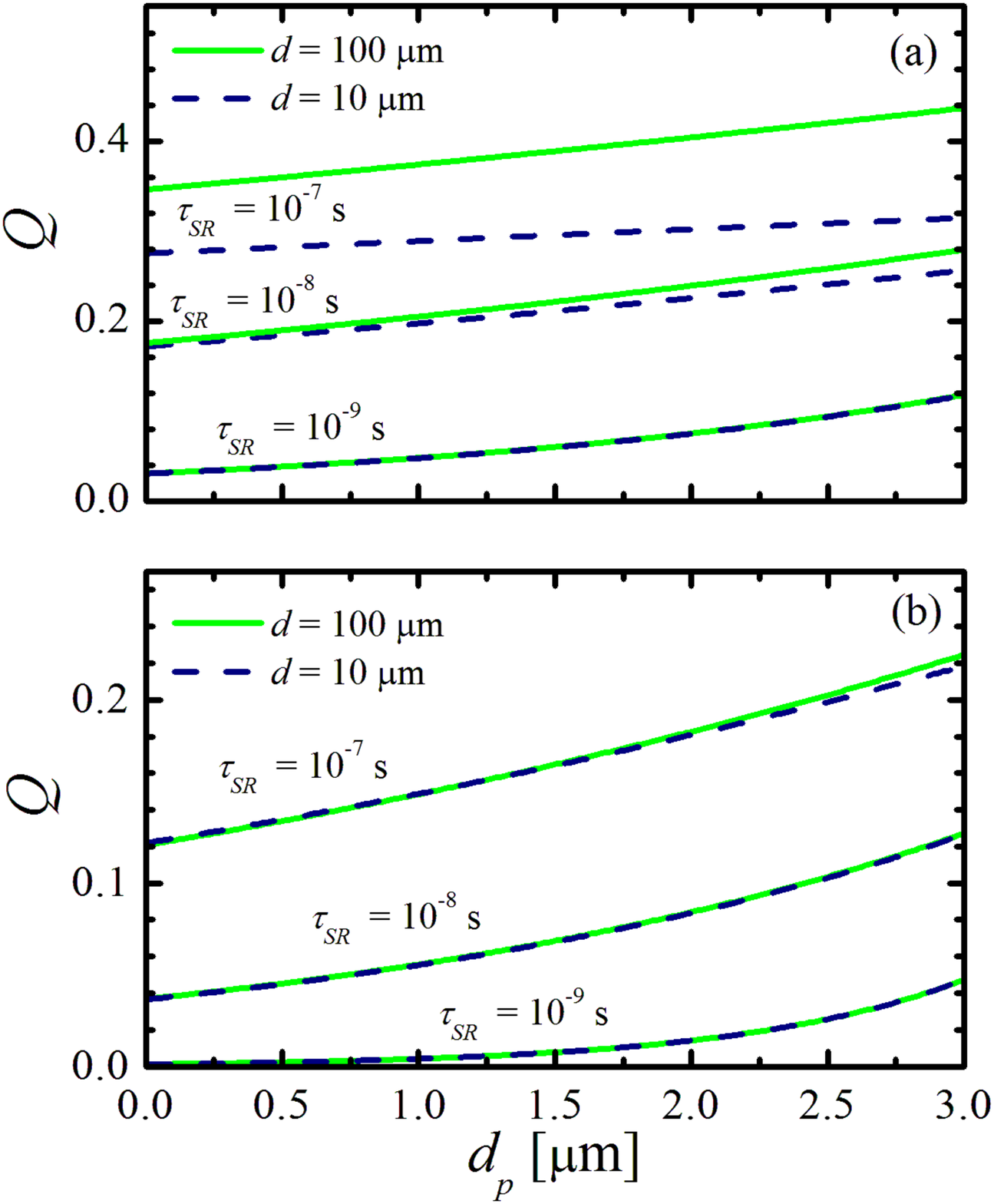}
\caption{Collection coefficient of a \Pm/GaAs betavoltaic element as a function of junction depth for different Shockley-Reed lifetimes and element thicknesses for the case of (a) p-type base and (b) n-type base. }
\label{fig4}
\end{figure}

Shown in Fig.~\ref{fig4} is the collection coefficient as a function of $d_p$ for a \Pm/GaAs pair calculated for different lifetimes $\tau_{SR}$ and junction thicknesses $d$ of 10 and 100\,$\mu$m. In this case, according to \cite{Dmi78}, $x_m = 3\cdot 10^{-4}$\,cm$^3$/s. Panels (a) and (b) correspond to the cases of p- and n-base conduction types, respectively. As seen in the figure, rather high values of $Q \ge 0.4$ for the \Pm/GaAs pair can be achieved only for the junction thickness $d \approx 100\,\mu$m. Also, collection coefficient decreases dramatically as $\tau_{SR}$ decreases.

It should be noted that similar results for the attainable $Q$ are expected for other direct-bandgap  A$_3$B$_5$ semiconductors, in particular, the ones based on the three-component compounds.

\section{Open-circuit voltage analysis}
When estimating the limiting efficiency value \cite{Olstech, Ada12}, we used the Shockley-Queisser approach \cite{Sho61}, in which not only the current density, but also the open-circuit voltage, $V_{OC}$, is assumed to be maximal. Therefore, our next task is to calculate the open-circuit voltage, $V_{OC}$, with realistic values of $\tau_{SR}$. It is given by the standard expression
\begin{equation}
V_{OC} = \frac{k_BT}{q}\ln\frac{N_d\Delta p^*}{n_i^2}\ ,
\label{11}
\end{equation}
where $\Delta p^* = \Delta p(x = d_p + w)$ is the excess minority carrier density in the base at the boundary between the space-charge region and quasilinear region of thickness $w$, $N_d$ is the equilibrium density of the majority carriers in the quasineutral base region, and $n_i$ is the intrinsic charge carrier density. It is related to the effective densities of states in the conduction and valence bands, $N_c$ and $N_v$, as
\begin{equation}
n_i = \sqrt{N_cN_v}\exp\left(-\frac{E_g}{2k_BT}\right)\ .
\label{17}
\end{equation}

We assume that both $d_p$ and $w$ are much smaller than the diffusion length $L$. This allows us to approximate 
\begin{equation}
\Delta p(x = 0) \approx \Delta p^*\ .
\end{equation} Such an approximation introduces a negligible error into $V_{OC}$ from Eq.~(\ref{11}) in view of its logarithmic dependence on $\Delta p^*$.

We will assume that recombination dominates in the quasineutral base region and in the space-charge region. Then, $V_{OC}$ can be obtained using the approach from \cite{Sach14}. Taking into account the generation-recombination processes, we first write the continuity equation for the excess carrier density supplemented by the boundary conditions:
\begin{eqnarray}
&&\frac{d^2\Delta p}{dx^2} - \frac{\Delta p}{L^2} - r(x)\,\Delta p(x)  + g(x) = 0\ , \nonumber \\
&&\frac{d\Delta p}{dx}(x = d) = 0\ ,\nonumber\\
&& D\frac{d\Delta p}{dx}(x = 0) = S_0\,\Delta p^*\ ,
\label{9}
\end{eqnarray}
where the third term describes recombination processes in the space-charge region of the abrupt junction, and the last one corresponds to the beta-induced generation. The first boundary condition is consistent with our assumption $S_d \ll D/L$ from the beginning of the previous section, and the second one is responsible for recombination effects in the $x = d_p + w$ plane.

Integration of the continuity equation results in the balance equation for the generation-recombination currents, according to which the current density for electronic excitation is proportional to the integral of the generation term,
\begin{equation}
J_\beta = q\,\int_0^d dx\,\frac{\Delta p(x)}{\tau_b} + q\left(S_0 + R_{SC}\right)\,\Delta p^*\ ,
\label{10}
\end{equation}
where $q$ is the elementary charge. The right-hand side in (\ref{10}) is responsible for the recombination in the bulk and on the front side of the emitter and within the space-charge region. The space-charge region recombination rate is given by \cite{Sze}
\begin{eqnarray}
&&R_{SC}(\Delta p^*) = \frac{L_D}{\sqrt{2}\tau_{SR}}\int_{y_{pn}}^{-1} dy\,N_d\,\left(1 - y + e^y\right)^{-1/2}\times \nonumber\\
&&\Big[N_d e^y + n_i e^{E_r/k_BT} + b\left(\frac{n_i^2}{N_d} + \Delta p^*\right)e^{-y} \nonumber\\
&&\ \ \ \ \ \ \ + b n_i e^{-E_r/k_BT}\Big]^{-1}\ ,
\nonumber
\end{eqnarray} 
where $b = \sigma_p/\sigma_n$ is the ratio of the capture cross-sections of holes and electrons by a recombination level, $E_r$ is the recombination level energy measured from the middle of the bandgap,  $y_{pn}$ is the dimensionless potential at the p-n boundary, $L_D$ is the Debye length.

To evaluate the first integral in (\ref{10}), we have employed the following approximative procedure. First, we write the solution of the continuity equation (\ref{9}) as a sum of homogeneous and inhomogeneous parts,
\begin{equation}
\Delta p(x) = \frac{e^{-x/L} + e^{(x-2d)/L}}{1 + e^{-2d/L}}\Delta p^* + \Delta p_i(x)\ ,
\end{equation}
where the homogeneous term satisfies the first boundary condition in (\ref{9}) and gives the value $\Delta p(x = 0) = \Delta p^*$. The inhomogeneous contribution $\Delta p_i(x)$, with $\Delta p_i(x = 0) = 0$, is notably different from zero only within a relatively thin layer below the front surface of the emitter, where the generation-recombination processes take place. Therefore, the contribution to the integral of the second term can be neglected in comparison to the integral of the homogeneous term, allowing us to write
\begin{equation}
\int_0^d dx\Delta p(x) \approx \Delta p^*L\tanh(d/L)\ .
\end{equation}
This approximation should produce a negligible error in $V_{OC}$ in view of its logarithmic dependence on $\Delta p^*$. Substitution of this result into Eq.~(\ref{11}) taking into account that $L^2 = D\tau_b$ yields
\begin{equation}
J_\beta = q\Delta p^*\left[\frac{D}{L}\tanh\left(\frac{d}{L}\right) +S_0 + R_{SC}(\Delta p^*)\right]\ .
\label{14}
\end{equation}

The current density $J_\beta$ is inversely proportional to the energy required to create one electron-hole pair, $\varepsilon$, which is approximately related to the bandgap $E_g$ as \cite{Klein68}
\begin{equation}
\varepsilon = 2.8\,E_g + 0.5\,\text{eV}\ .
\label{10a}
\end{equation}
Denoting is the current density in the case of Si ($E_g = 1.12$\,eV) by $J_0$, the current density in the case of arbitrary bandgap can be approximated as 
\begin{equation}
J_\beta =J_0\,Q\cdot 3.64\,\text{eV}/\varepsilon\ .
\end{equation}
We note that, usually, $J_0$ is in the $1$ -- $10^2\,\mu$A/cm$^2$ range \cite{Olstech}. The value of $\Delta p^*$ found from Eq.~(\ref{14}) should be substituted into Eq.~(\ref{11}) to obtain the open-circuit voltage $V_{OC}$.

\begin{figure}[t!] 
\includegraphics[scale=0.25]{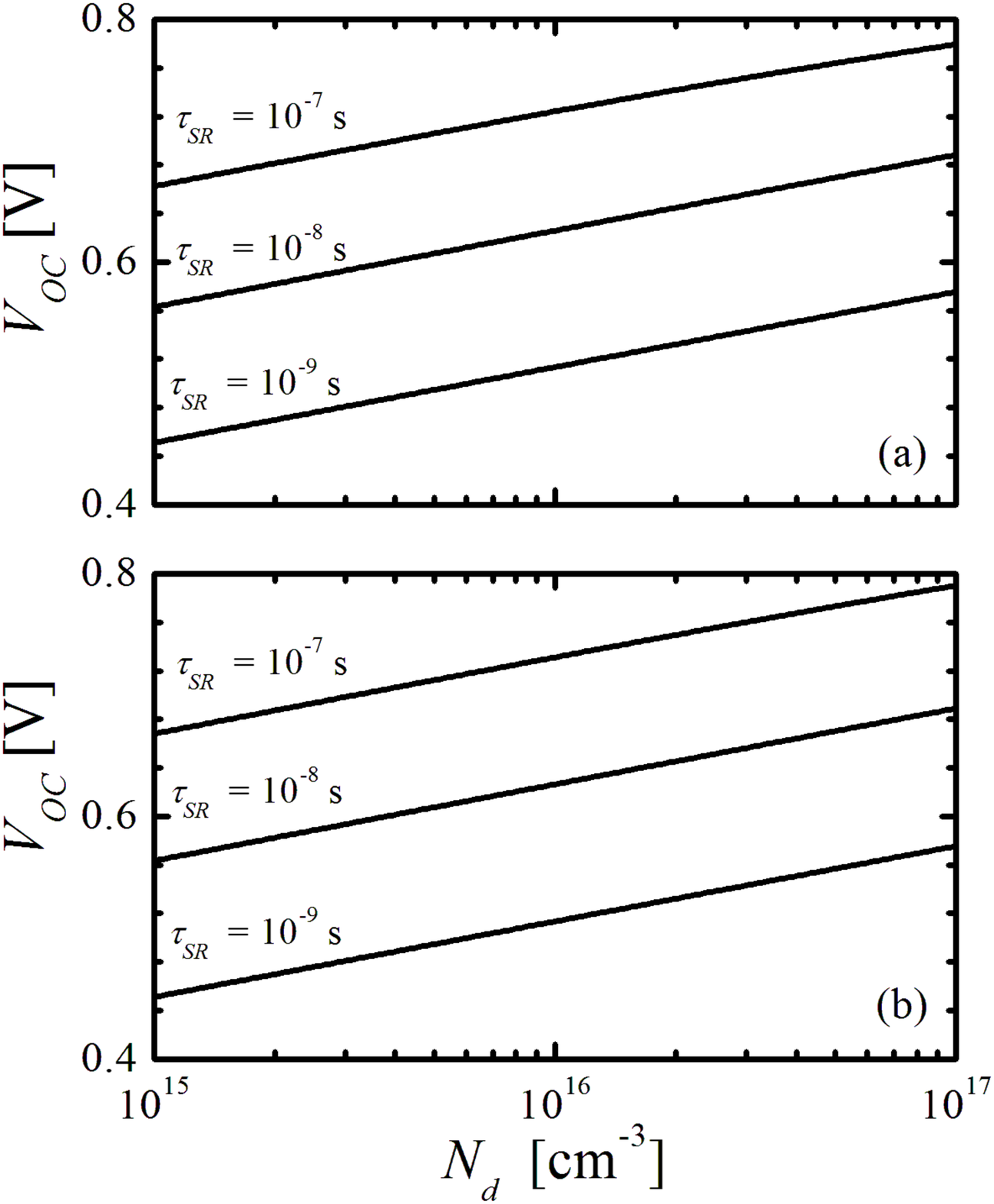}
\caption{Open-circuit voltage as a function of the base doping level for different Shockley-Reed lifetimes in the case of (a) p-type base and (b) n-type base for $E_g$ = 1.43\,eV, $T = 300$\,K, and $J_0 = 10$\,$\mu$A/cm$^2$.}
\label{fig5}
\end{figure}

Fig.~\ref{fig5} shows the dependence of $V_{OC}$ of a GaAs-based p-n junction on the base doping level, $N_d$, neglecting the surface recombination, that is, $S_0 \approx 0$. As seen in Fig.~\ref{fig5}, $V_{OC}$ increases with $N_d$. On the one hand, the values of $V_{OC}$ for the pair \Tr/GaAs is notably smaller than in the solar cells \cite{Sach14}, because the beta-produced current densities are at least two order of magnitude smaller than the short-circuit current densities in photovoltaic cells. On the other hand, the open-circuit voltages in Fig.~\ref{fig5} exceed the values obtained experimentally in \cite{And00}. The reason is that, in \cite{And00}, the current density $J_0$ was of the order of $1\,\mu$A/cm$^2$, whereas in our calculations, we have taken $J_0 = 10\,\mu$A/cm$^2$. If the values $J_0 = 1\,\mu$A/cm$^2$, $N_d = 5\cdot 10^{16}$\,cm$^{-3}$, and $\tau_{SR} = 10^{-9}$\,s are used, we obtain $V_{OC} = 0.44$\,V, which practically coincides with the value given in \cite{And00}.

\section{Refined calculation of the limiting betaconversion efficiency}
According to Olsen \cite{Olstech}, the efficiency of a betavoltaic element, $\eta$, is
\begin{equation}
\eta = \eta_\beta\,\eta_C\,\eta_S\ ,
\end{equation}
where
\begin{equation}
\eta_\beta = N_\beta/N_0
\end{equation}
is the fraction of beta-flux that reaches the semiconductor,
\begin{equation}
\eta_C = (1 - r)\,Q
\end{equation}
is the coupling efficiency, given by the product of absorption probability of a beta-particle ($r$ is the electron reflection coefficient from the semiconductor surface) and collection efficiency $Q$ of electron-hole pairs, and, finally, the semiconductor efficiency
\begin{equation}
\eta_S = q\,V_{OC}\,FF/\varepsilon\ ,
\end{equation}
where $q$ is the elementary charge, $V_{OC}$ is the open-circuit voltage, $FF$ is the fill factor, $\varepsilon$ is the energy necessary to generate one electron-hole pair from Eq.~(\ref{10a}).

Let us obtain $V_{OC}$ within the Shockley-Queisser approximation, where $\tau_{SR} \to \infty$, $S_0 and R_{SC} \to 0$, and the only recombination mechanism present is radiative recombination, characterised by the coefficient $A$. In this case, $V_{OC}^{lim}$ can be found analytically from (\ref{10}) and (\ref{14}):
\begin{equation}
V_{OC}^{lim} = \frac{k_BT}{q}\ln\frac{J_\beta}{qAdn_i^2}\ , \\
\label{16}
\end{equation}

The fill factor can be found using the expression from \cite{Olstech}
\begin{equation}
FF = \left[v_{OC} - \ln(v_{OC} + 0.72)\right]/(v_{OC} + 1)\ ,
\label{18}
\end{equation}
where $v_{OC} = V_{OC}/k_BT$.

To calculate the limiting beta-conversion efficiency, we take $Q = 1$, $r = 0$, $\eta_\beta = 1$, corresponding to the bidirecional source in the terminology of \cite{Olstech}. In this case
\begin{equation}
\eta_{lim} = \frac{q\,V_{OC}^{lim}\,FF_{lim}}{2.8\,E_g + 0.5}\ ,
\label{19}
\end{equation}
where $V_{OC}^{lim}$ is given by (\ref{16}). 

When calculating $\eta_{lim}$, several issues may arise. First, the parameters $N_c$, $N_v$, and $A$ are material-specific in every semiconductor. Second, when evaluating $V_{OC}^{lim}$ and $FF_{lim}$, Olsen had used, for each source, concrete current density $J_0$ of the order of $10^2\,\mu$A/cm$^2$ for \Pm\ and $1\,\mu$A/cm$^2$ for \Tr. Finally, $V_{OC}^{lim}$ depends on the p-n junction thickness $d$. Therefore, all parameters in (\ref{19}) must be specified. Since such key parameters as $A$, $N_c$, and $N_v$ are known only for concrete semiconductors and concrete bandgap values $E_g$, in the best-case scenario, the dependence $\eta_{lim}(E_g)$ can be found as a set of support points for the known semiconductors with different $E_g$. Fitting this with a smooth curve might not be accurate enough.

In this work, we calculated $\eta_{lim}$ only for the case of GaAs using Eq.~(\ref{19}). For $A = 2\cdot 10^{-10}\,cm^3$/s and $d = 10\,\mu$m gives for $J_0 = 10^2\,\mu$A/cm$^2$ the value $\eta_{lim} \approx 17$\,\%, and for $J_0 = 1\,\mu$A/cm$^2$, $\eta_{lim} \approx 14$\,\%. Note that the values of $\eta_{lim}$ obtained here notably exceed the ones obtained by Olsen in \cite{Olstech, Ols12}. In the rest of this work, we will use the values obtained for the \Pm/GaAs and \Tr/GaAs combinations, respectively.

\begin{figure}[t!] 
\includegraphics[scale=0.25]{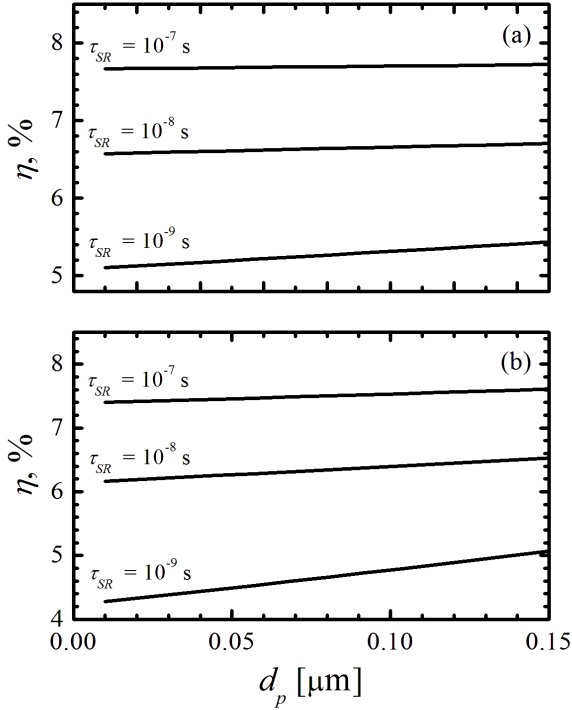}
\caption{Beta-conversion efficiency of a \Tr/GaAs pair as a function of junction depth for different Schokley-Reed lifetimes for the case of (a) p-type base and (b) n-type base.}
\label{fig6}
\end{figure}

\begin{figure}[t!] 
\includegraphics[scale=0.25]{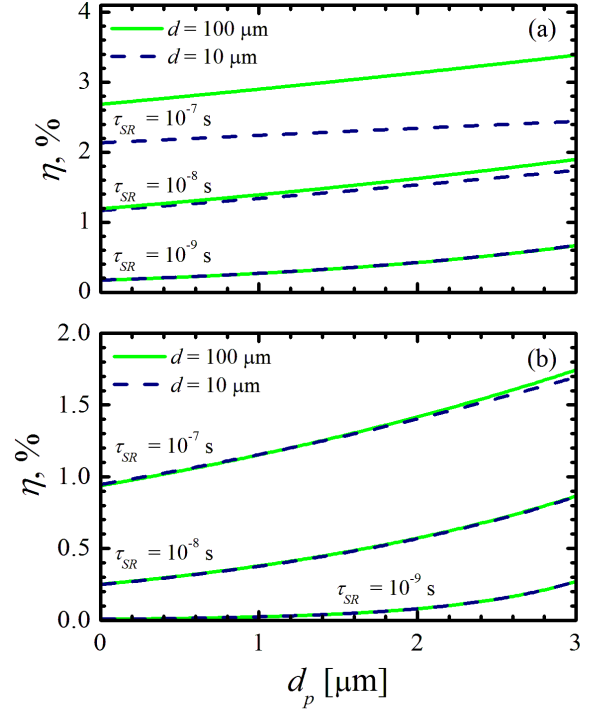}
\caption{Beta-conversion efficiency of a \Pm/GaAs element vs. junction depth for different Schokley-Reed lifetimes and element thcknesses for (a) p-type base and (b) n-type base.}
\label{fig7}
\end{figure}

\section{Calculation of the attainable betaconversion efficiency}
Fig.~\ref{fig6} shows the attainable efficiency as a function of $d_p$ for the \Tr/GaAs combination, obtained from
\begin{equation}
\eta = \eta_{lim}Q\frac{V_{OC}}{V^{lim}_{OC}}\ ,
\label{20}
\end{equation}
where $\eta_{lim} \approx 14$\,\%, $Q$ is given by Eq.~(\ref{8}), $V_{OC}$ is found from Eq.~(\ref{11}), and $V_{OC}^{lim}$  from Eq.~(\ref{16}).

When plotting Fig.~\ref{fig6}, we varied the lifetime at a constant $d = 10\,\mu$m. Panels (a) and (b) correspond to the base of the p- and n-type, respectively. As seen in Fig.~\ref{fig6}, the attainable efficiency values are rather high and are in the range of (6.4 - 12.5)\%.

It should be noted that our results for \Tr/GaAs pair agree well with those given in the review \cite{Ols12} citing Refs.~\cite{And00, Bow02, Ada12}, namely, $\eta =$ (4 - 7) \%. In these works, a \Tr-source was used with the $A_3B_5$-based semiconductors. But, as evident from the figures shown, the possibilities of increasing the efficiency of \Tr/$A_3B_5$ betaconversion are far from being exhausted.

Shown in Fig.~\ref{fig7} is the attainable beta-efficiency (\ref{20}) as a function of $d_p$ for \Pm/GaAs pair with $\eta_{lim}$ = 17\,\%. The $\tau_{SR}$ values used were $10^{-9}, 10^{-8}$, and $10^{-7}$\,s, and GaAs thicknesses were 10 and 100 $\mu$m. Fig.~\ref{fig7}(a) and (b) correspond to the p- and n-types of the base conductivity. As seen in this figure, $\eta$ reduces rather strongly as $\tau_{SR}$ is decreased. For the highest $\tau_{SR} = 10^{-7}$\,s, $\eta$ decreases with decreasing $d$. The highest efficiency attainable, $\eta = 7.25$\,\%, is achieved for $\tau_{SR} = 10^{-7}$ s and $d = 100\,\mu$m, and the lowest value of $0.51$ \% is realized for $\tau_{SR} = 10^{-9}$ s and $d = 10\,\mu$m.

Thus, we conclude that a \Pm/GaAs-based betaconverter is not as efficient as a \Tr/GaAs-based one. Perhaps, the very small efficiency of the \Pm/GaAs battery obtained in \cite{Fli64} is due to the small thickness of GaAs and small lifiteme $\tau_{SR}$. The same applies also to the cases when, instead of GaAs, other direct-bandgap semiconductors are used.

\section{Conclusions}
Our analysis, focusing on the attainable collection coefficient $Q$ and open-circuit voltage values $V_{OC}$, has revealed the following features of current collection of the GaAs-based beta-elements.

Efficient collection of the electron-hole pairs generated by a beta-flux can be achieved when the diffusion length exceeds the dead layer thickness, $L > x_m$. An alternative way to increase collection coefficient is to use deep junctions, for which $d_p \simeq x_m$.

Additional mechanisms responsible for the reduction of current generated by beta-electrons are possible, leading to smaller betaconversion efficiency. They may be due, for instance, to the strong absorption of the beta-electrons by auxiliary layers of a betavoltaic element.

Using the Shockley-Queisser approximation, we have derived the limiting betaconversion efficiency, $\eta_{lim}(E_g)$. Our analysis has shown that, because the main parameters affecting the efficiency are very different for different semiconductors, the $\eta_{lim}(E_g)$ curve can be build as a set of support points for semiconductors with different bandgaps, and not as a smooth curve.

\Pm\ beta-source performs more poorly than \Tr-source, because the electron-hole pair generation depth in the case of \Pm-source is large, whereas the diffusion length of GaAs is small. Therefore, the majority of electron-hole pairs generated in the base recombine before reaching the p-n junction.

In the case of \Tr-source, the picture is different. The collection coefficient is rather high, because of the small generation depth of electron-hole pairs. Therefore, the realistic betaconversion efficiency for the \Tr/GaAs pair will be rather high for relevant parameters (lifitemes and diffusion coefficients) of the semiconductor.

Similar results are expected also in the case, when other direct-bandgap semiconductors are used instead of GaAs.

\acknowledgments
M.E. would like to thank Natural Sciences and Engineering Research Council of Canada (NSERC) for financial support.

\end{document}